\documentclass[fleqn,twoside]{article}
\usepackage{espcrc2}

\usepackage{graphicx}
\usepackage[figuresright]{rotating}

\newcommand{\AmS}{{\protect\the\textfont2
  A\kern-.1667em\lower.5ex\hbox{M}\kern-.125emS}}

\title{Status report on  {\tt TAUOLA},
       its environment, and its applications}

\author{Z. W\c as\address[MCSD]{Institute of Nuclear Physics,
         Radzikowsiego 152 , 31-342 Cracow, Poland.\\ and \\
         CERN, Theory Division, CH-1211 Geneva 23, Switzerland.}%
        \thanks{
This work is partly supported by
the Polish State Committee for Scientific Research 
(KBN) grants 2 P03B 001 22, 
and also by the European Community's Human Potential
Programme under contract HPRN-CT-2000-00149 Physics at Colliders. }
 \thanks{Home page at http://wasm.home.cern.ch/wasm/},
}
       
\begin{document}

\begin{abstract}

The status of the Monte Carlo programs for the simulation of the $\tau$-lepton production 
and decay in 
high energy accelerator experiments  is reviewed. In particular, the status of the
following packages is discussed: (i) {\tt TAUOLA} for $\tau$-lepton decay and  {\tt PHOTOS} for 
radiative corrections in decays,  (ii) MC-TESTER for universal tests of the Monte Carlo programs
describing particle decays, (iii) {\tt KORALB, KORALZ, KKMC} packages for $\tau$-pair production 
in $e^+e^-$ collisions, and (iv)  universal
interface of {\tt TAUOLA} for the decay of $\tau$-leptons produced by ``any'' generator.
\vspace{1pc}

\hspace {28.5pc} {\bf CERN-TH/2002-302}
\end{abstract}

\maketitle

\section{INTRODUCTION}
The  package {\tt TAUOLA}
\cite{Jadach:1990mz,Jezabek:1991qp,Jadach:1993hs,Golonka:2000iu}  for the simulation 
of $\tau$-lepton decays and  
{\tt PHOTOS} \cite{Barberio:1990ms,Barberio:1994qi} for the simulation of radiative corrections
in decays, have a rather long history. Written and maintained by 
well defined authors, they nonetheless migrated into a wide range
of applications, where they became ingredients of 
complicated simulation chains. As a consequence, a large number of
different versions are currently in use.
From the algorithmic point of view, they often
differ only in a few small details, but incorporate
substantial amounts of specific results from distinct
$\tau$-lepton measurements. Such versions were mainly maintained  
by the experiments taking precision data on $\tau$-leptons. On the other hand,
 many new applications were developed  recently,  often requiring
 program different versions  because of
interfaces to other packages.

In the following, I will concentrate on  those topics where changes  with respect
to the status presented at the Victoria $\tau$ conference two years ago \cite{Was:2000st}
were introduced.
Since that time, there were no changes introduced into the {\tt PHOTOS} Monte Carlo
functionality, and also the {\tt TAUOLA} interfaces to {\tt KORALB} 
\cite{koralb2:1995,jadach-was:1984},
{\tt KORALZ} \cite{koralz4:1994},  and {\tt KKMC}  \cite{kkcpc:1999}
remain unchanged. On the contrary, the universal interface of {\tt TAUOLA} evolved, 
and new applications,
in particular observables for the measurement of the Higgs-boson parity,  
became possible. The new program {MC-TESTER} \cite{Golonka:2002rz} instrumental 
in the development 
of future versions of 
{\tt TAUOLA}, was developed. A new  choice of hadronic 
currents for $\tau \to 4 \pi \nu$ decay modes became available.
It is based on Novosibirsk data.

Let me concentrate, in the following three sections, on these topics, 
 and close my contribution with a summary.

\section{THE $\tau$-LEPTON FOR HIGGS BOSON PARITY}

In many applications, the precision of the control of the $\tau$ spin effects is not crucial.
This is the case in  searches of new particles yet to be discovered,
or in applications where $\tau$-lepton decays contribute as a final state of some rare
decays of known particles. Nonetheless, in such cases, spin effects can be of some use 
as well. As these, there is no motivation to develop sophisticated spin algorithms 
for every individual case: less precise, but universal solutions are welcomed.

The
universal solution of \cite{Pierzchala:2001gc}, based on the {\tt  HEPEVT} common block 
of {\tt FORTRAN77} is now distributed  with {\tt TAUOLA}.  The basic idea was
to calculate the spin (helicity) state of the decaying $\tau$ from the 
kinematical information available in {\tt  HEPEVT} and some very simple assumptions
on the production mechanisms. The program was checking if the production was through 
$ f \bar f \to Z/\gamma \to \tau^+ \tau^-$,  $W$ or Higgs-boson intermediate states/processes. 
The main properties of the algorithm were  presented in Victoria already 2 years ago;
modifications necessary for the program to work in the case 
of some $\tau$ production processes involving new, to-be-discovered particles, have 
gradually been introduced, but one can see \cite{Bisc} that not in all cases
is the effort completed.

 In the meantime algorithm functionality was also extended: full spin effects were introduced
in the case 
of Higgs-boson decay \cite{Was:2002gv}. 
The case of the Higgs boson is exceptionally easy, since the
full density matrix of the $\tau$-lepton pair produced from a Higgs-boson  is
fully defined  by boson parity and $\tau$ leptons four-momenta. 

This algorithm helped us to design an observable
\cite{Bower:2002zx} 
for the possible measurement of Higgs boson parity in future accelerators such as
Linear Colliders or the LHC.  Let us recall the main principle of this technique, which from 
the point of view of spin analysis is quite involved, as it requires a 
study of full spin correlation spanning over {\it three} levels of decay cascade:
$h/A \to \tau^+ \tau^-$, $\tau^\pm \to \rho^\pm \nu$ and $\rho^\pm \to \pi^\pm \pi^0$.
The presence of two non-observable neutrinos as well as our inability to reconstruct the 
Higgs-boson rest frame sufficiently well (with a precision comparable to the  
$\tau$-lepton mass) complicates the picture even further. 

\begin{figure}[htb]
\vspace{-4pc}
\includegraphics[height=20pc,width=20pc]{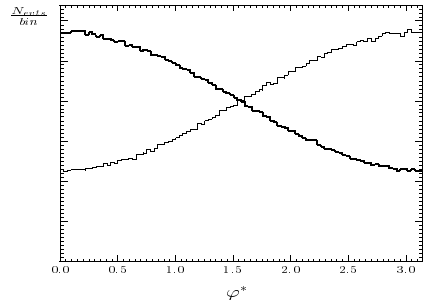}
\vspace{-9pc}
\caption{\it Acoplanarity distribution of the $\rho^+ \rho^-$ decay products' 
in the rest frame of the $\rho^+ \rho^-$ pair. A cut on   
 the $\pi^\pm$ to $\pi^0$ energy difference defined, in the generator 
level $\tau^\pm$ rest frames, to be of  the same sign is used. No smearings included.
 The thick line denotes the case   
of the scalar Higgs-boson and the thin line the pseudoscalar one. }
\label{MClevel}
\vspace{-2pc}
\end{figure}

We  started from the observation
that the distribution of the acoplanarity angle of two planes spanned on 
decay products of $\rho^+ \to \pi^+ \pi^0$ and    $\rho^- \to \pi^- \pi^0$
(defined in the rest frame of a $\rho^+\rho^-$ pair) is quite sensitive, see fig. \ref{MClevel},
to the parity of Higgs boson. One has to  select sub-samples of particles,
 with  the same sign (opposite sign) of the  energy difference between 
 $\pi^+$ and $ \pi^0$  in the $\tau^+$ rest frame  and that between
 $\pi^-$ and $ \pi^0$ in the $\tau^-$ rest frame. 
This condition is unfortunately impossible to realize in practice, as
$\tau$-lepton momenta cannot be reconstructed. 

\begin{figure}[htb]
\vspace{-4pc}
\includegraphics[height=20pc,width=20pc]{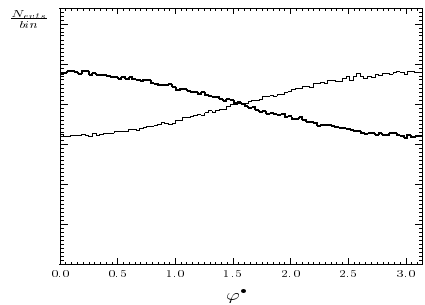}
\vspace{-9pc}
\caption{\it Acoplanarity distribution of the $\rho^+ \rho^-$ decay products'     
in the rest frame of the $\rho^+ \rho^-$ pair. A cut on the  
differences of the $\pi^\pm$ to $\pi^0$ energies defined, in their respective replacement  
$\tau^\pm$ rest frames, to be of  the same sign   
is used. All detector smearings are included.  
 The thick line denotes the case   
of the scalar Higgs boson and the thin line the pseudoscalar one.  }
\label{smear}
\vspace{-2pc}
\end{figure}

In the next step of our study we have weakened the requirement, and 
we have used a replacement for the $\tau$-lepton momentum; the
$\rho^+$ and $ \rho^-$ directions (in the rest frame of a $\rho^+\rho^-$ pair)
replace the  unobservable directions of $\tau$ flights and the   
Higgs mass constraint provides  the $\tau$-lepton energy (also in the same frame).
Indeed, with such a technique, and after including some assumption on detector smearings,
the sensitivity of our observable diminished significantly (see fig. \ref{smear}),
but nonetheless remained sizeable.
Enough to guarantee the meaurement of Higgs boson parity to a confidence level greater than 
95\%, if typical assumptions on linear collider luminosity, and  (120 GeV)
Higgs-boson production mechanism are made. 
Our method proved quite stable, with a degrading assumption on angular 
and energy resolution of the detector. Even a reduction by a factor of 2--3  
with respect to typical assumptions used in linear collider designs did 
not destroy the method. To preserve its sensitivity, the angular resolution must be 
$\le {\cal O }(m_\rho/E_\rho)$.

\section{ MC-TESTER}

In the previous case the {\tt HEPEVT} data structure was used 
to search for possible hard processes to calculate the spin state of the decaying $\tau$.
One can search over an event record such as  {\tt HEPEVT} for other information as well.
An essential step before attempting any sizeable rebuilding of the {\tt TAUOLA} library
 is to automate at least some
of its tests (such rebuilding may soon become necessary because of improving $\tau$-lepton data).
Such a test-program may be useful for other applications, not necessarily related 
to generators of $\tau$-lepton decay, as well. 
That is why we have developed an individual program for that
purpose. The idea behind
{\tt MC-TESTER} \cite{Golonka:2002rz} is quite simple: 
the user  loads an extra library 
and, after the generation of every event with his generator, calls {\tt MC-TESTER},  specifying the
identifier for the particle to be searched for over the whole event record and studied. 
A data file from such a run is  formed.

\begin{figure}[htb]
\vspace{-4pc}
\includegraphics[height=40pc,width=25pc]{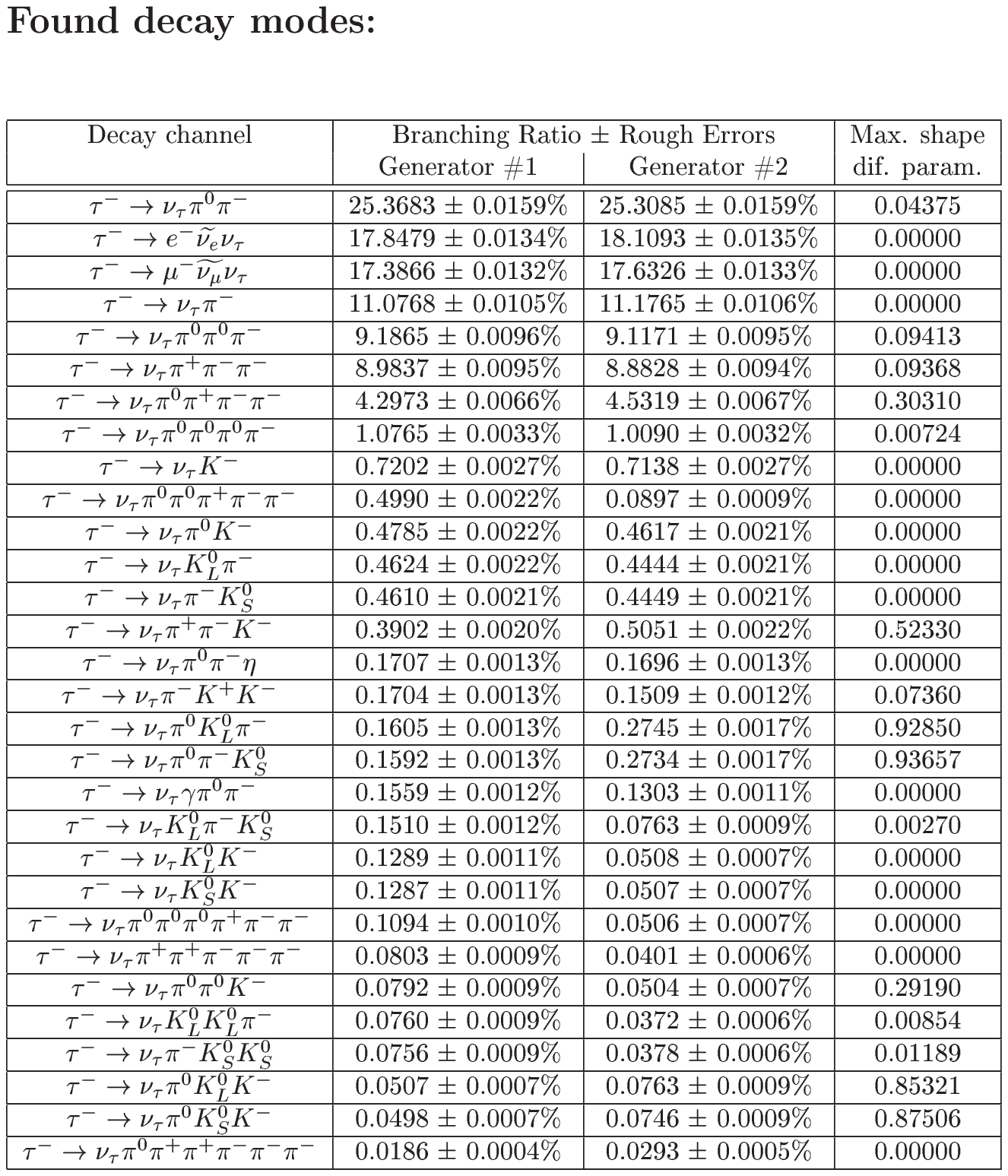}
\vspace{-9pc}
\caption{\it Second page of {\tt MC-TESTER} booklet produced  at analysis step.}
\label{table}
\vspace{-2pc}
\end{figure}

\begin{figure}[htb]
\vspace{-4pc}
\includegraphics[height=20pc,width=18pc]{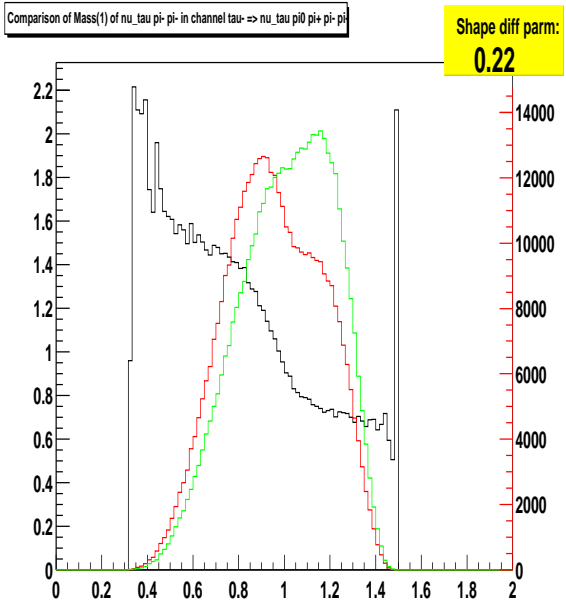}
\vspace{-1pc}
\caption{\it Example of individual plot from analysis step of {\tt MC-TESTER}.}
\label{shapediff}
\vspace{-2pc}
\end{figure}

In the second (analysis) step, data files from  two runs of different Monte Carlo programs
 can be compared. The
output is given in the form of a \LaTeX{} file, which  includes a
 table of all  decay modes found(see fig. \ref{table}). 
The table is not only a list of these modes,
but also, of the branching ratios calculated from the two runs.
Number quantifying maximum of shape differences is also given.  This number is  calculated
 as a maximum
of the shape difference parameters (with a part of the code easy to identify)
for all invariant
mass distributions of the decay channel under study. We choose to take into consideration
invariant mass of  every subset of the decay products 
for the decay channel.
The table is followed by a booklet of these invariant mass distributions  (see  
fig. \ref{shapediff} for an example of an individual plot), which are grouped
 into separate chapters for every decay mode.

\section{{ TAUOLA}}
As  was already discussed in Victoria  \cite{Was:2000st}, many options of physics 
initialization are now available for {\tt TAUOLA} (they can be constructed
from the files included  in the program distribution package).
They often differ by the models used for fits
to the experimental data, and/or by the  data themselves.
Let me stress that, in general, the best parametrizations 
will always be in the hands of the leading $\tau$-lepton experiments of 
the  time. Our original version of {\tt TAUOLA} initialization
was indeed meant to be used only after re-initialization of its physics content by
the users. Nonetheless for   experiments
in the design phase or for which $\tau$-leptons  physical processes help 
in other studies, such a solution is not good. We are  gradually including 
new options into {\tt TAUOLA}. 

In the last two years, parametrizations of the hadronic 
form factors \cite{Bondar:1999ac} based on the Novosibirsk 
 low energy $e^+e^-$ annihilation data  \cite{Akhmetshin:1998df} 
were introduced into {\tt TAUOLA} as 
a possible new option for $\tau^- \to \pi^-\pi^- \pi^+ \pi^0 \nu$ and 
$\tau^- \to \pi^-\pi^0 \pi^0 \pi^0 \nu $ decay modes \cite{Bondar:2002mw}.
This is the only published contribution documenting upgrades on the physics
content of {\tt TAUOLA} over the last two years.

\subsection{Comment on one of the old tests}

In ref. \cite{Jadach:1993hs}  (formulae  (37) and (38))
we have reported that, for $\tau$ decay into
4$ \pi $, when the chiral  limit is taken (the case we used for some tests
of phase-space generation), the results of {\tt TAUOLA} agree with those of
ref. \cite{Fischer:1979fh},  {\it only} up to an overall factor%
\footnote{
Let us note that problems with normalization (or/and with misprints) in 
\cite{Fischer:1979fh} were again addressed 
 \cite{Ecker:2002qi}. The authors of that paper worry 
that it may 
be of grave consequences for {\tt TAUOLA} as well. In practice, 
even if the normalization of the chiral current was wrong
(by the same factor in the Monte Carlo and in the testing function), 
it would  affect neither technical tests nor actual
results of {\tt TAUOLA} generation, since a different current is used then.
 Note that
in cases such as {\tt CLEO} initialization \cite{cleo}, the current 
is normalized
directly from the $\tau$ data collected by the experiments; the normalization
originating from the model disappears completely.
This is why the issue is only of 
marginal interest for {\tt TAUOLA} as it is now.}
${243 \over 2 \cdot n!} \cos^2 \theta_c $.
In the case of $\tau \to \pi^- \pi^- \pi^+ \pi^0 \nu$, ($n=2$) the  test formula 
\begin{equation}
{\Gamma(2\pi^- \pi^+ \pi^0) \over \Gamma_e} =  { \cos^2\theta_c\over 15 }
\Bigl({m_\tau \over 2\pi f_\pi}\Bigr)^4{1 \over 128} C_N
\end{equation}
 returns 0.0271364 with the numerical coefficient $C_N=1261/120 -\pi^2$
(for compatibility, we use as in \cite{Jadach:1993hs} $m_\tau=1.7842$, $f_\pi=0.0933$ and 
$\cos \theta_c=0.975$). 
A later calculation (formula 33 in \cite{Czyz:2000wh}) finds 
$C_N=1009/96 -\pi^2$  and result is 0.0272249. 
With the computer power available at the time of \cite{Jadach:1993hs}, 
we could not distinguish between the two options.
At present, such an exercise is straightforward and we find that {\tt TAUOLA} returns
$0.0272060 \pm 0.000011$. We can see, that the value of $C_N$
from ref.~\cite{Czyz:2000wh} 
is statistically favoured over the one from ref.~\cite{Fischer:1979fh}.

\section{{ SUMMARY}}

Let me recall finally the main developments of the {\tt TAUOLA} package in the last two years: 
as stand-alone generator of $\tau$-lepton decays, 
and for its interfaces to other programs.
The progress on $\tau$-lepton decay alone consists
of a new parametrization of form factors for the $4\pi$ decay channels. 
This parametrization, mainly based on the Novosibirsk data, form a step toward 
{\tt TAUOLA}  as a framework, where different data and models can be compared.
 In such multi parametrization approach the appropriate tools are necessary. 
The {\tt MC-TESTER}, helpful in different types  of works for  programs such 
as {\tt TAUOLA} became available. It  can also be useful, for example,  to check if 
the interfacing 
of {\tt TAUOLA} to other programs did not corrupt its initialization.
The family  of {\tt TAUOLA}  interfaces  was enriched with the new option for its
universal interface. It now provides complete spin correlations for 
Higgs boson decay into pair of $\tau$ leptons. This option 
was instrumental in designing an observable that is  potentially useful to measure Higgs-boson 
parity at future  experiments.

Let me finally acknowledge  people and collaborations
who  contributed to the present shape of {\tt TAUOLA} and related projects:
A. Bondar, S. Eidelman, P. Golonka, S. Jadach, M. Je\.zabek, J.H.~K\"uhn,
A. Milstein,  T. Pierzchala,  E. Richter-Was, N. Root, B.F.L. Ward, M. Worek, and 
ALEPH, CLEO (in particular A. Weinstein), Delphi, Opal, L3 collaborations, the 
Karlsruhe theory group, and others. I would like to thank A. Pich for discussion.


\begingroup\begin{thebibliography}{10}

\bibitem{Jadach:1990mz}
S.~Jadach, J.~H. K\"uhn and Z.~W\c as, {\em Comput. Phys. Commun.} {\bf 64} (1990)
275.

\bibitem{Jezabek:1991qp}
M.~Je\.zabek, Z.~W\c as, S.~Jadach and J.~H. K\"uhn, {\em Comput. Phys. Commun.} {\bf
  70} (1992)
69.

\bibitem{Jadach:1993hs}
S.~Jadach, Z.~W\c as, R.~Decker and J.~H. K\"uhn, {\em Comput. Phys. Commun.} {\bf
  76} (1993)
361--380.

\bibitem{Golonka:2000iu}
P.~Golonka, E.~Richter-W\c as and Z.~W\c as,
\href{http://www.arXiv.org/abs/hep-ph/0009302}{{\tt hep-ph/0009302}}.

\bibitem{Barberio:1990ms}
E.~Barberio, B.~van Eijk and Z.~W\c as, {\em Comput. Phys. Commun.} {\bf 66}
  (1991)
115.

\bibitem{Barberio:1994qi}
E.~Barberio and Z.~W\c as, {\em Comput. Phys. Commun.} {\bf 79} (1994)
291--308.

\bibitem{Was:2000st}
Z.~W\c as, {\em Nucl. Phys. Proc. Suppl.} {\bf 98} (2001) 96--102,
\href{http://arXiv.org/abs/hep-ph/0011305}{{\tt hep-ph/0011305}}.

\bibitem{koralb2:1995}
S.~Jadach and Z.~W\c as, {\em Comput. Phys. Commun.} {\bf 85} (1995) 453--462.

\bibitem{jadach-was:1984}
S.~Jadach and Z.~W\c{a}s, {\em Acta Phys. Polon.} {\bf B15} (1984) 1151;
  \uppercase{E}rratum: {\bf B16} (1985) 483.

\bibitem{koralz4:1994}
S.~Jadach, B.~F.~L. Ward and Z.~W\c{a}s, {\em Comput. Phys. Commun.} {\bf 79}
  (1994) 503.

\bibitem{kkcpc:1999}
S.~Jadach, B.~F.~L. Ward and Z.~W\c{a}s, {\em Comput. Phys. Commun.} {\bf 130}
  (2000) 260; an up-to-date source is available from http://home.cern.ch/jadach/.

\bibitem{Golonka:2002rz}
P.~Golonka, T.~Pierzcha\l a and Z.~W\c as,
\href{http://arXiv.org/abs/hep-ph/0210252}{{\tt hep-ph/0210252}}.

\bibitem{Pierzchala:2001gc}
T.~Pierzcha\l a, E.~Richter-W\c as, Z.~W\c as and M.~Worek, {\em Acta Phys. Polon.}
  {\bf B32} (2001) 1277--1296,
\href{http://arXiv.org/abs/hep-ph/0101311}{{\tt hep-ph/0101311}}.

\bibitem{Bisc}
C.~Biscarat, web page at http: //home.cern.ch/cbiscara/tauola$\_$valid.html.

\bibitem{Was:2002gv}
Z.~W\c as and M.~Worek, {\em Acta Phys. Polon.} {\bf B33} (2002) 1875--1884,
\href{http://arXiv.org/abs/hep-ph/0202007}{{\tt hep-ph/0202007}}.

\bibitem{Bower:2002zx}
G.~R. Bower, T.~Pierzcha\l a, Z.~W\c as and M.~Worek, {\em Phys. Lett.} {\bf B543}
  (2002) 227--234,
\href{http://arXiv.org/abs/hep-ph/0204292}{{\tt hep-ph/0204292}}.

\bibitem{Bondar:1999ac}
A.~E. Bondar, S.~I. Eidelman, A.~I. Milstein and N.~I. Root, {\em Phys. Lett.}
  {\bf B466} (1999) 403--407,
\href{http://arXiv.org/abs/hep-ph/9907283}{{\tt hep-ph/9907283}}.

\bibitem{Akhmetshin:1998df}
{CMD2} Collaboration, R.~R. Akhmetshin {\em et al.}, {\em Phys. Lett.} {\bf
  B466} (1999) 392--402,
\href{http://arXiv.org/abs/hep-ex/9904024}{{\tt hep-ex/9904024}}.

\bibitem{Bondar:2002mw}
A.~E. Bondar {\em et al.}, {\em Comput. Phys. Commun.} {\bf 146} (2002)
  139--153,
\href{http://arXiv.org/abs/hep-ph/0201149}{{\tt hep-ph/0201149}}.

\bibitem{Fischer:1979fh}
R.~Fischer, J.~Wess and F.~Wagner, {\em Z. Phys.} {\bf C3} (1980)
313--320.

\bibitem{Ecker:2002qi}
G.~Ecker and R.~Unterdorfer,
\href{http://arXiv.org/abs/hep-ph/0209056}{{\tt hep- ph/0209056}}.

\bibitem{cleo}
{CLEO} Collaboration, A.~Weinstein, see 
  http: //www.cithep.caltech.edu/\~{}ajw/korb$\_$doc. html\#{}files.

\bibitem{Czyz:2000wh}
H.~Czy\.z and J.~H. K\"uhn, {\em Eur. Phys. J.} {\bf C18} (2001) 497--509,
\href{http://arXiv.org/abs/hep-ph/0008262}{{\tt hep-ph/0008262}}.

\end{thebibliography}\endgroup

\providecommand{\href}[2]{#2}

\end{document}